# 3D Aeronomy Modeling of Close-in Exoplanets

I. F. Shaikhislamov[1], M. L. Khodachenko[2,3], H. Lammer[2], A. G. Berezutsky[1],
I. B. Miroshnichenko[1], M. S. Rumenskikh[1]

*1) Institute of Laser Physics SB RAS, Novosibirsk, Russia*
*2) Space Research Institute, Austrian Academy of Sciences, Graz, Austria*
*3) Skobeltsyn Institute of Nuclear Physics, Moscow State University, Moscow, Russia*
*E-mail address: ildars@ngs.ru*

**ABSTRACT**

We present a 3D fully self-consistent multi-fluid hydrodynamic aeronomy model to study the structure of a hydrogen dominated expanding upper atmosphere around the hot Jupiter HD 209458b and the warm Neptune GJ 436b. In comparison to previous studies with 1D and 2D models, the present work finds such 3D features as zonal flows in upper atmosphere reaching up to 1 km/s, the tilting of the planetary outflow by Coriolis force by up to 45 degrees and its compression around equatorial plane by tidal forces. We also investigated in details the influence of Helium ($He$) on the structure of the thermosphere. It is found that by decrease of the barometric scale-height, the $He$ presence in the atmosphere strongly affects the $H_2$ dissociation front and the temperature maximum.

## 1. INTRODUCTION

Development of gas-dynamic models to simulate expanding upper atmospheres of close-in orbiting exoplanets has been steadily progressing since the discovery of first hot Jupiters in early 2000s. It was realized, as early as in *Lammer et al.* (*2003*) that due to the high stellar XUV fluxes their upper atmospheres should be in a hydrodynamic regime where in some cases the outflow may result even in super-sonic blow off. The tantalizing detection of absorption in the Ly$\alpha$ line at transit of HD 209458b (*Vidal-Madjar et al. 2003, 2004*) exceeding the optical transit depth by several times gave strong support of such scenarios with a formation of hydrodynamically expanding upper atmosphere, that fills the Roche lobe of the planet. Since the first 1D numerical simulation by (*Yelle et al. 2004*) a number of works have been carried out, for example, by *Tian et al. (2005)* and *Guo (2011)*. Besides of hydrogen and *He*, other relevant species, such as nitrogen, oxygen and carbon (*N, O, C*), were simulated in *García Muñoz (2007)* and *Koskinen at al. (2007)*. The effect of tidal force was studied by *Erkaev et al. (2007)* and *Trammell et al. (2011)*. *Murray-Clay et al. (2009)* have shown that Ly$\alpha$ cooling is important, as well as the reabsorption of Ly$\alpha$ photons (*Shaikhislamov et al. 2014*), while in *Yelle et al.* (*2004*), *Koskinen at al.* (*2013a*) and *Shaikhislamov et al.* (*2014*) it has been demonstrated that the planetary atmosphere escape is also affected by cooling via $H_3^+$ molecule infrared emission. All those previous studies have been carried out with 1D hydrodynamic models, whereas some important features of the problem require 2D and 3D modeling. In particular, the tidal forces in a planet-centered reference frame can be accurately modeled only in 3D. The interaction of the expanding planetary wind (PW) with the stellar wind (SW) requires at least a 2D modeling, while in the case of close-orbit planets with substantial Keplerian velocities, the full 3D simulations are needed. The interaction between SW and PW appears to be important for the interpretation of transit observations, because it is responsible for generation of Energetic Neutral Atoms (ENAs) (*Holmström et al. 2008*) and for shaping of the planetary plasma environment (*Shaikhislamov et al. 2016, Khodachenko et al. 2017*). The first 2D or 3D models, applied to exoplanets (*Stone and Proga 2009*), have been adopted from the available astrophysical codes and did not include the physics of the PW formation, which was prescribed as a fixed boundary condition. However, these models revealed a number of important features unavailable in 1D, such as shock formation in the colliding PW and SW (*Stone and Proga 2009*), different regimes of planetary atmospheric material escape (*Matsakos et al. 2015; Shaikhislamov et al. 2016*), its global structuring affected by the SW (*Bisikalo et al. 2013; Matsakos et al. 2015*), and generation of ENAs (*Tremblin and Chiang 2013*). A possible magnetic field of a planet adds another aspect to the problem, which can be only approximately modeled in 2D (*Trammell et al. 2014; Khodachenko et al. 2015*), but in case of the inclusion of the SW, it requires a full 3D simulation. If the stellar magnetic field, frozen-in the SW, is sufficiently strong, its treatment also requires a 3D modeling (*Erkaev et al. 2017*). Besides of a hydrodynamic approach, a 3D Monte-Carlo code is used for the modeling of exoplanetary upper atmospheres and exospheres as well. It allows tracing of escaping planetary particles and calculating the ENAs, generated by charge-exchange with the SW protons and due to acceleration of neutral hydrogen atoms by the stellar radiation (*Vidal-Madjar et al. 2003; Lecavelier des Etangs et al. 2008; Bourrier et al. 2013, 2015; Kislyakova et al. 2014*).

Despite of the extensive use of numerical simulations, and new observational results, a self-consistent model that would combine the physics of PW formation (starting from dense and warm planetary atmosphere) with its global 3D dynamics in the hot SW plasma environment, encompassing a whole system, has not been developed yet. The 3D hydrodynamic codes used so far remain severely limited. For example, in *Tripathi et al.* (*2015*) a simplifying assumption consists in taking a monochromatic XUV flux and consideration of the initial atmosphere, consisting of atomic hydrogen only, without inclusion of molecular component. Their results depend also on the prescribed hydrogen density at the inner boundary, which indicate on the incomplete account of the absorbed stellar XUV radiation energy. In *Matsakos et al.* (*2015*) and *Bisikalo et al.* (*2013*) the PW flow is not calculated, but rather launched by prescribing particular values as the boundary conditions, that correspond to the thermosphere temperature maximum, positioned at a height of $\sim 0.5 R_p$ above the photometric radius.

In this study we present a multi-fluid 3D model, which self-consistently describes the generation of the expanding PW, energized by the absorbed stellar XUV radiation, and its large-scale propagation away from the planet. The model includes the hydrogen aeronomy and photo-chemistry, as well as the wavelength dependent radiation transfer. It has been evolved from our previous 1D (*Shaikhislamov et al. 2014; Khodachenko et al. 2015*) and 2D (*Shaikhislamov et al. 2016; Khodachenko et al. 2017*) models and retains all the features included therein. An important difference consists in 3D geometry, which enables treating of all forces (e.g., the tidal force and Coriolis force) without any approximation. The planet-centered spherical coordinate system with an exponentially varying radial mesh allows sufficiently detailed resolution of a highly stratified planetary atmosphere and covers at the same time the whole of the planet's orbit including the host star.

In the present study we use this 3D model to calculate a planetary wind relatively close to a planet assuming that it is unaffected by stellar plasma. The aim is to see how 3D effects shape the escaping particle flow close to the planet in the regions of the upper thermosphere and the exosphere, as well as outside the Roche lobe. In that sense this study is a follow up of the previous 1D aeronomy modeling of close-orbit exoplanets, such as in *García Muñoz* (*2007*), *Koskinen et al.* (*2013a*), *Shaikhislamov et al.* (*2014*) where the SW was not taken into account. The problem as a whole is quite complex and before presenting full 3D simulation with the PW interacting with the SW in a global setting including a star it is useful to make a preliminary step and present results concerning only the PW formation. As mentioned before, another goal is to investigate how the $He$ abundance affects the planetary wind. While the previous studies included $He$ as component of atmosphere, its influence to the outward flowing PW has not been investigated. This, however, might be of interest for several reasons. First, because of intensive mass loss the ratio $He/H$ might greatly vary between exoplanets during their lifetime, especially for low-mass close-in orbiting planets, being a kind of an indicator of the planetary evolutional path (*Hu et al. 2015*). Second, even for the expected proto-planetary abundance of $He/H=0.1$ the $He$ constitutes more than 40% of mass, and therefore it appears an important agent in the planetary mass loss. Third, $He$ is much more efficient absorber of XUV radiation below 50 nm than hydrogen, thus its presence affects the heating of atmosphere. In support of the mentioned above importance of the account of $He$ in the models, we note of recent and first actual observational confirmation of presence of $He$ around the hot Jupiter WASP-107b (*Spake et al. 2018*).

As particular objects for the 3D modeling, demonstrated in this paper, we take one of the most studied exoplanets, a hot Jupiter HD 209458b and another intriguing planet, warm Neptune GJ 436b. The first object remains of great interest because of the still controversial and incomplete interpretation of transit measurements in Ly$\alpha$ and in resonant lines of carbon, oxygen and some other species (*Vidal-Majar et al. 2004; 2008, 2013; Ben Jaffel and Sona Hosseini 2010; Koskinen et al. 2013b*). Our recent study with a 2D code revealed that tidal acceleration of planetary streams is crucial for the explanation of thermal line broadening absorption by $OI$, $CII$ and $SiIII$ (*Shaikhislamov et al. 2018*). However, an accurate quantitative treatment of this effect requires a 3D modeling of the material motion around the planet inside and outside the Roche lobe. Another object investigated here, GJ 436b, shows an extremely high absorption in blue wing of the Ly$\alpha$ line (*Ehrenreich et al. 2015, Lavie et al. 2017*) that requires a consistent interpretation. The recent 1D aeronomy modeling of GJ 436b by (*Loyd at al. 2017*) revealed that its exosphere is weekly ionized and extends rather far from the planet. However, it cannot produce a blue Doppler-shifted absorption in velocity range of 50–120 km/s. The velocity of the expanding PW is below 10 km/s, which is to be expected for the relatively low radiation flux of the M-dwarf GJ436, whereas the claim made on the basis of Monte-Carlo simulations (*Burrier at al. 2015*), that the absorption in Ly$\alpha$ can be explained if the planetary wind is accelerated by Alfven waves to velocities of 70 km/s, remains hypothetical. Therefore, further investigation of the dynamical near-by plasma environment of GJ 436b is an actual task.

The paper is organized in the following way. In Section 2 we briefly describe the model; in Section 3 the results of simulations, performed for HD 209458b and GJ 436b are presented, and Section 4 is dedicated to the discussion and conclusions.

**2 THE MODEL**

The details of our numerical model have been described in (*Shaikhislamov et al. 2014, 2016*) and in (*Khodachenko et al. 2015, 2017*). Here we provide just a short description of the most important characteristics of the code and highlight the new features related with 3D geometry. The plasma is treated as a quasi-neutral fluid

(due to the ambipolar condition), which is in a partial thermal equilibrium, i.e. with equal electron and ion temperatures, while that of neutral components might be different. As the initial state of simulations, we take a fully neutral atmosphere in a barometric equilibrium composed of $H_2$ and neutral $He$ at different mixing ratios. The mixing ratio is defined as relation of Helium to Hydrogen nuclei $\chi = He/H$. Thus, the content of $He$ in initial atmosphere is given by $n_{He} = 2\chi \cdot n_{H_2}$. As the first approximation of a steady state we take that all species have the same scale height based on mean molecular mass. After that the actual distribution is calculated in the model self-consistently according to acting forces and ion-atomic collisions. However, the model doesn't include such processes as molecular and eddy diffusion which are likely to bias the $H_2$/He redistribution at pressures higher than ~1 µbar. The eddy diffusion requires an extensive modeling itself and in aeronomy codes is treated empirically (for example, *García Muñoz 2007*).

We employ a multi-fluid approach, which deals with $H_2, H, H^+, He, He^+$ species. For more realistic treatment of the thermosphere, the model takes into account also the $H_2^+, H_3^+$ molecular ions. The list of modeled hydrogen reactions is presented in *Khodachenko et al.* (*2015*), and it is practically the same as that used in other aeronomy models (e.g., *García Muñoz 2007; Koskinen et al. 2007*), except that no reactions between *He* and *H* are considered. At the conventional surface of the planet, which is an inner boundary of the simulation domain, a fixed temperature $T_{base}$, pressure $P_{base}$ and mixing ratio are set. The boundary is chosen sufficiently deep in the atmosphere, typically at a pressure of $P_{base}=0.03$ bar, so that all XUV radiation with a modeled spectrum is completely absorbed. This ensures that the whole energy of the stellar XUV radiation flux is deposited to the atmosphere and the simulation results regarding its expansion and material escape do not depend on the position of the inner boundary (*Shaikhislamov et al 2014*). It should be noted that at such dense atmosphere the composition and chemistry of atmosphere is strongly influenced by other elements such as *N*, *C*, *O*. However, as involved aeronomy codes show (*García Muñoz* 2007), the parameters of atmosphere above $1.1 R_p$ is affected by it only in a little degree.

As the characteristic values of numerical problem, we use the radius of a planet $R_p$ (different for different planets considered below), the temperature $10^4$ K, and the corresponding thermal velocity of protons $V_o=9.07$ km/s, so that the obtained numerical solutions are correspondingly scaled in these units. To achieve sufficient spatial resolution in a highly stratified atmosphere, a radial step close to the planet surface is taken as small as at least $\Delta r=R_p/400$. The typical azimuthal and polar angle steps are of about $4°$. For visualization of the simulation results, we use Cartesian coordinate system with *X*-axis pointing to the star, and *Y*-axis opposite to the orbital velocity.

The photo-ionization due to the stellar XUV, which results in heating by photo-electrons and increase of the pressure gradient, is the main driver of the PW. For the modeling of HD 209458b, the Solar XUV flux is assumed that covers 10–912 Å spectral range, binned at 1 Å steps (*Tobiska 1993*). The spectrum is based on measurements of the solar radiation under the conditions of a moderate activity index with $P_{10.7}=148$. The integrated XUV flux with this spectrum at 1 AU is $F_{XUV}=4.466$ erg s$^{-1}$ cm$^{-2}$. To account for slightly larger size of HD 209458, we take as a base level the flux value of $F_{XUV}=6$ erg s$^{-1}$ cm$^{-2}$ at 1 AU. The XUV photons ionize hydrogen atoms, as well as $H_2$ and $He$. The model assumes that the energy released in the form of photo-electrons is rapidly and equally re-distributed between all locally present particles with an efficiency of $\eta_h = 0.5$. The attenuation of the XUV flux inside the planetary atmosphere is calculated for each spectral bin according to the wavelength dependent cross-sections. For the M-dwarf GJ 436 we use the XUV spectrum compiled by the MUSCLES survey (*France et al. 2016*). Because of the significantly smaller stellar size, its integrated XUV flux is only of about $F_{XUV}=0.85$ erg s$^{-1}$ cm$^{-2}$ at 1 AU.

Other processes, affecting the particle distributions and mutual transformations are electron impact excitation and ionization, dielectronic recombination, charge-exchange and elastic collisions. For the typical parameters of the planetary exospheres, Coulomb collisions with protons effectively couple all ions. For example, at $T<10^4$ K and $n_{H^+} > 10^6$ cm$^{-3}$ the collisional equalization time for the temperature and momentum $\tau_{Coll} \approx 10^6 T^2/(n_{H^+} V_{H^+}) \cdot (M_i/m_p)$ (*Braginskii, 1965*) is less than 1 s for protons. This is several orders of magnitude less than the typical gasdynamic time scale of the problem $R_p/V_o$, which is of the order of $10^4$ s. Therefore, there is no need to calculate the dynamics of every charged component of the plasma, and we assume in the simulations all of them (i.e., $H^+, H_2^+, H_3^+$, $He^+$) to have the same temperature and velocity. On the other hand, the temperature and velocity of each neutral component are calculated individually.

The tidal potential of the system in the rotating, planet-centred frame of reference, which can be found, for example, in *Trammell et al. (2011)*, includes the planetary and the stellar gravity, as well as the centrifugal force due to the planet rotation. The latter in the case of a tidally locked planet is synchronized with its orbital revolution. Besides of that, the Coriolis force $\sim 2m\vec{V} \times \vec{\Omega}$ is added to the momentum equations.

In the present modeling we don't include the SW plasma flow and restrict the calculations by distances around the planet, closer than $10R_p$. While SW affects the global structure of planetary exosphere (see e.g., in *Shaikhislamov et al. 2016*), it does not influence significantly the processes responsible for the upper atmosphere expansion and formation of PW, such as the radiative heating and proto-chemical transformations, which take place sufficiently close to the planet. We therefore assume the SW to be weak enough, not to affect the inner exosphere at the considered close-in distances. This simplification is of course a rather specific, but not unrealistic one.

The physical (relevant) parameters of the modeled stellar-planetary systems were taken from http://www.exoplanet.eu, and are summarized in Table 1.

**Table 1.** Parameters of exoplanets used in simulations.

|  | HD 209458b | GJ 436b |
|---|---|---|
| mass | $0.71 M_J$ | $0.07 M_J$ |
| radius | $1.38 R_J$ | $0.35 R_J$ |
| orbit radius | 0.047 AU | 0.029 AU |
| period | 3.52 days | 2.64 days |
| star mass | $1.148 M_{Sun}$ | $0.45 M_{Sun}$ |

## 3 HD 209458b

Figure 1 demonstrates the spatial structure of the nearby plasma environment around HD 209458b, calculated for $F_{XUV}=6$ erg s$^{-1}$ cm$^{-2}$ at 1 AU, with $T_{base}=1250$ K, $P_{base}=0.03$ bar, and $He/H=0.05$, which we will consider further as standard ones for this planet.

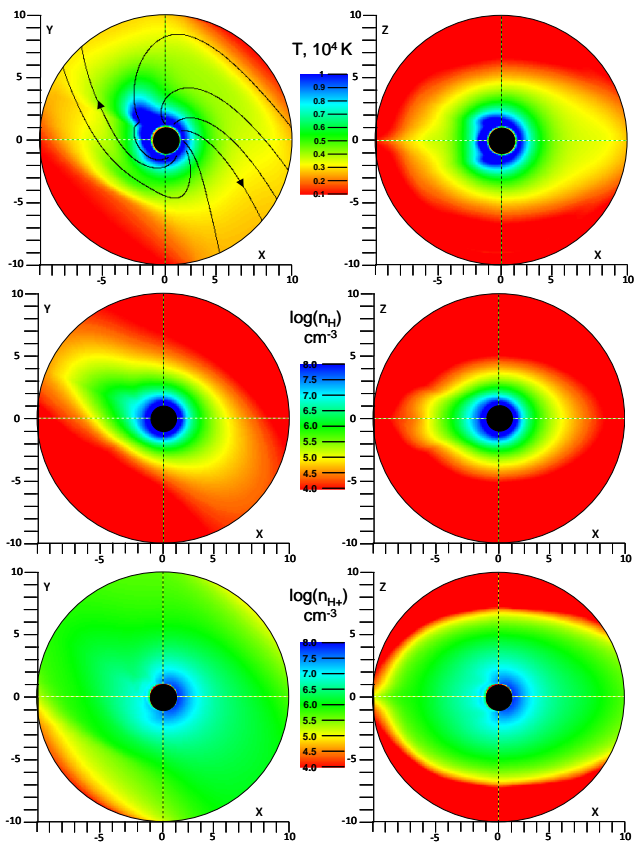

**Figure 1.** Distribution of temperature (upper row) and density of hydrogen atoms (middle row), as well as protons (bottom row) in the equatorial $X$-$Y$ (left column) and the meridional $X$-$Z$ (right column) planes, respectively, calculated for HD209458b with $F_{XUV}=6$ erg s$^{-1}$ cm$^{-2}$ at 1 AU, $T_{base}=1250$ K, $P_{base}=0.03$ bar, $He/H=0.05$. The distance is scaled in units of HD209458b radius, $R_p$. The star is located at the right at $X=70 R_p$. The first panel also shows material velocity streamlines. Black circle with $r=R_p$ at the center marks the planet.

The streamlines show that in a tidally locked reference frame the material is rotated clockwise by inertia forces as it flows away from the planet. Correspondingly, the temperature and the density distributions in the equatorial plane are twisted as well. In our previous 2D simulations (*Khodachenko et al., 2015, Shaikhislamov et al. 2016*) the temperature maximum, due to zonal flows, was observed at the nightside. In 3D, due to the twisting, it appears close to the terminator as well. The atomic hydrogen density decreases away from the planet because of the photo-ionization and the plasmasphere extends up to ~$10 R_p$.

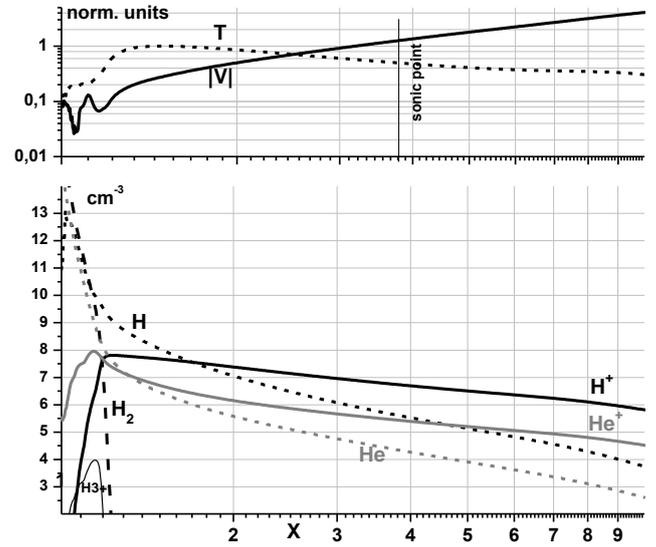

**Figure 2A.** Profiles of temperature and absolute velocity (upper panels, in units of $10^4$ K and $V_o$) and density of various components (bottom panels, log scale, in cm$^{-3}$, dashed black – $H_2$, dotted black – $H$, solid black – $H^+$, dashed grey – $He$, solid grey – $He^+$, thin black – $H_3^+$) along the planet-star line. Parameters of simulations are the same as in Fig. 1 except that simulation box is $20 R_p$.

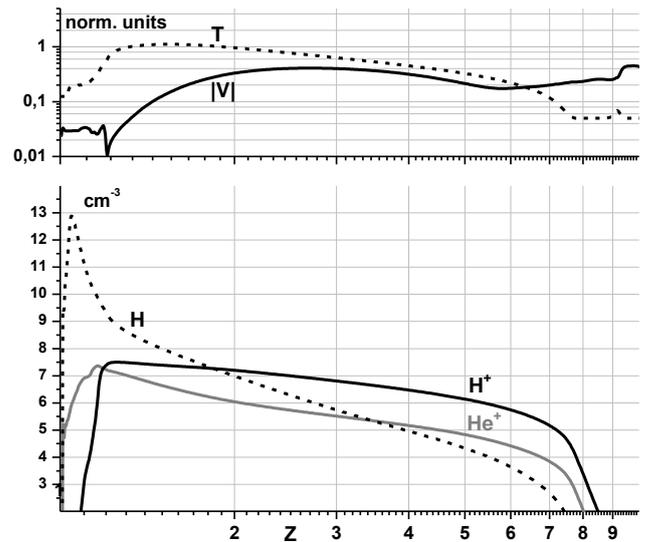

**Figure 2B.** Profiles of temperature and absolute velocity (upper panels) and density of various components (bottom panels, log scale, in cm$^{-3}$, dotted black – $H$, solid black – $H^+$, solid grey – $He^+$,) along Z-axis. Parameters of simulations are the same as in Fig. 1 except that simulation box is $20 R_p$.

The protons, at the same time, spread in the equatorial plane more or less homogenously. In the meridian plane, one can see a sharp boundary in the protons distribution, which was also observed in our previous 2D simulations (*Khodachenko et al. 2015*). This meridional inhomogeneity is formed because of the stellar gravity pull, which has above the poles a component towards the equatorial plane, not compensated by the centrifugal force (*Trammell et al. 2011*). Despite this effect the thermal pressure driving the atmosphere is strong enough so the planetary wind carries material more or less uniformly across the whole Roche boundary.

Detailed information on the structure of simulated nearby plasma environment around HD209458b is given in the graphs in Figure 2. In particular, because of the efficient cooling due to the infrared emission by $H_3^+$ ions and dissociation of $H_2$, the temperature varies only slightly up to the height of about $1.25R_p$. At these altitudes the azimuthal zonal flow dominates in the material motion (see in upper left panel). Between $1.25R_p$ and $1.5R_p$ the temperature increases rapidly to its maximum of about $10^4$ K, and the PW material starts its expansion away from the planet. Above $1.5R_p$ velocity continues to increase, passing super-sonic point at the distance of about $3.9R_p$ along the planet-star line, while temperature adiabatically decreases. Density profiles show that hydrogen molecules and the chemistry processes, in which they are involved, exist up to the height of about $1.3R_p$. Above that, only $H$, $H^+$, $He$, $He^+$ are present. Half-dissociation point ($n_H=n_{H2}$) is located at a pressure level of about ~1 μbar, while half-ionization point ($n_{H+}=n_H$) at a height of about $1.8R_p$. The profiles along the polar axis reveal a sharp boundary at about $8R_p$ beyond which the PW material doesn't penetrate. It has been specially checked with a series of dedicated simulation runs, that this is an effect of stellar gravity, and the obtained position of the boundary well agrees with our previous 2D simulations (*Khodachenko et al. 2015*).

It should be noted that in the present 3D simulation, like in the previous 2D one (*Shaikhislamov et al. 2018*), the region of the dominating molecular hydrogen population extends farther from the planet surface (up to $1.25R_p$) as compared to the results of our previous works (*Shaikhislamov et al. 2014, Khodachenko et al. 2015*) as well as to other aeronomy models (*García Muñoz 2007, Koskinen et al. 2010*), where this region was limited by heights of about $1.1R_p$. A dedicated investigation of this result reveals that it is related with the cooling effect of $H_2$ dissociation via the reaction $H_2+H \rightarrow 3H$. Another influencing factor is that for $H_3^+$ cooling we use the emissivity calculated by *Miller et al.* (*2013*), which is several times larger than the one, used in the above mentioned works. Note that because of $H_2$ dissociation and $H_3^+$ cooling, the temperature below $1.25R_p$, despite of the absorption of XUV energy, practically does not increase. Without $H_2$ dissociation cooling and with a reduced $H_3^+$ emissivity, we obtain the same maximum height of $H_2$ presence as in our previous works (*Shaikhislamov et al. 2014, Khodachenko et al. 2015*). It should be noted that the rates of some important reactions are still not measured at temperatures of interest and are based on the extrapolations or theoretical calculations. Considering these differences, the results of modeling of even a relatively simple hydrogen dominated atmosphere can be corrected in a future studies by at least a factor of two.

The mass loss rate (further abbreviated as $\dot{M}$) is one of the main characteristics of the escaping PW. We calculate it as the material flux across the sphere co-centered with the planet. Because of the mass conservation it doesn't depend on the exact radius of the sphere across which the flux is evaluated. For example, in our simulations the difference between $\dot{M}$ calculated at $5R_p$ and $9R_p$ is 0.4%. Figure 3 shows how quickly the simulated PW of HD209458b reaches the steady state after switching the XUV flux on, for different abundances of helium. One can see that $\dot{M}$ saturates after 150-200 characteristic times $R_p/V_o$.

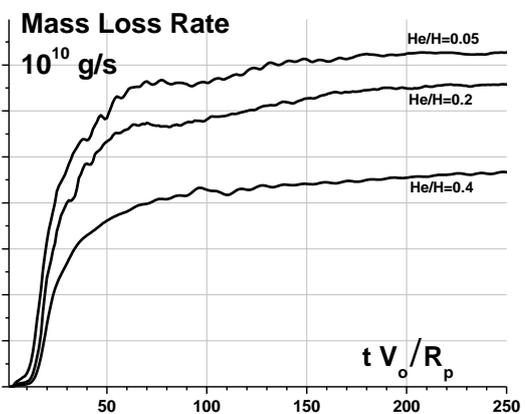

**Figure 3.** Dependence of mass loss rate calculated through the sphere $r=9R_p$ in dependence on simulation time for different Helium abundances in initial atmosphere.

At moderate $He$ content the value of $\dot{M}$ amounts to about $7.3 \cdot 10^{10}$ g/s. This is very close to our previous calculations made with 2D simulations, $\dot{M} \approx 7 \cdot 10^{10}$ g/s. When the total weight of $He$ in atmosphere exceeds that of hydrogen (for mixing ratio $He/H=0.4$), the $\dot{M}$ perceptibly decreases. Besides of that, $He$ affects significantly the temperature and velocity of the escaping PW and due to that, the size of the molecular hydrogen dominated region close to the planet. An important influence of helium consists also in its contribution to the mean barometric gravitational scale height, which decreases in inverse proportion to the mean molecular mass $\overline{m}_i$. The larger $\overline{m}_i$, the faster density of atmosphere decreases with height at the same temperature. By this, faster decrease of $H_2$ density with height results in a contraction of molecular dominated region. This is demonstrated in profiles of $H_2$ density and temperature in the simulated upper thermosphere, obtained for small ($He/H=0.01$), moderate ($He/H=0.05$), and large ($He/H=0.4$) helium abundances (see in, Figure 4). One can see that as soon as $H_2$ density falls below ~ $10^{12} - 10^{11}$ cm$^{-3}$ the temperature starts to increase due to

XUV heating, because the endothermic dissociation of $H_2$ cannot restrict it anymore. This in turn leads to progressive $H_2$ extinction.

Let's estimate the dissociation cooling and XUV heating at the dissociation front. The rate of $H_2+H \to 3H$ reaction depends exponentially on temperature with corresponding cooling rate of:

$$R_{diss} = 1.4 \cdot 10^{-8} e^{-5.5/T} \cdot n_{H1} \cdot n_{H2} \cdot T^{-1} \cdot E_{diss} \quad (1)$$

(UMIST Database for Astrochemistry, here $T$ is in units of $10^4$ K). The XUV heating rate $R_{XUV}$ at this layer exponentially depends on integrated column density of absorbing particles $NL = 2\sigma_o \int n_{H2} dl$, $\sigma_o = 6.3 \cdot 10^{-18}$ cm$^2$ is photo-ionization cross-section at threshold. The dependence $R_{XUV}(NL)$ can be found by integrating over the given XUV spectra, and the empirical approximate relation is

$$R_{XUV} \approx 10^{-16} \cdot (1+0.2NL)^{-1.2} \text{ erg cm}^{-3} \text{ s}^{-1} \quad (2)$$

(Trammell et al. 2011). The line integral $NL$ can be approximated by barometric length-scale assuming static equilibrium: $\int n_{H2} dl \approx n_{H2} \cdot H$, where for HD 209458b $H \approx T/\overline{m_i} \cdot 10^9$ cm. Thus, $NL = 6.3 \cdot 10^{-9} T \cdot n_{H2}$. From energy balance $R_{diss} \approx R_{XUV}$ we get $R_{XUV}/NL \approx 2 \cdot e^{-5.5/T} \cdot T^{-2} \cdot (n_{H1}/n_{H2}) E_{diss}$. Assuming $T=0.2$ and $n_{H1}/n_{H2} = (0.1-1)$ this gives $R_{XUV}/NL \approx (0.3-3) \cdot 10^{-22}$ erg cm$^{-3}$ s$^{-1}$. Finally, it follows that critical height is at $NL \approx (0.7-2) \cdot 10^3$ corresponding to $n_{H2} \approx (0.5-1.5) \cdot 10^{12}$ cm$^{-3}$. This order of magnitude estimate verifies that the observed in simulations correlation of the position of $H_2$ dissociation front and the height of thermosphere where temperature starts to rise with the layer where $H_2$ density falls below a certain level is indeed related to energy balance of dissociation cooling and XUV heating.

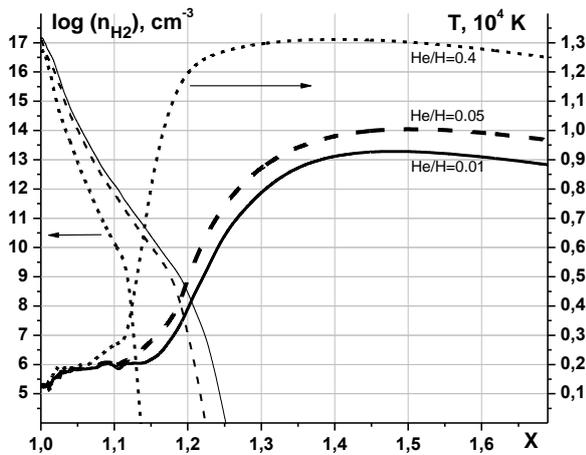

**Figure 4.** Profiles of density of molecular hydrogen and temperature along the planet-star line for different Helium abundances (He/H=0.01 solid, He/H=0.05 dashed, He/H=0.4 dotted).

In Figure 4 one can see also that the larger He content leads to higher maximum temperatures. This is because the heavier gas expands slower and in a shorter range of distances, resulting in less adiabatic cooling which influences the value of the maximum temperature in the thermosphere. Thus, the presence of helium in the upper atmosphere distinctly influences the position of the dissociation front and thermosphere low boundary where the temperature starts to increase, as well as the position and the value of the temperature maximum.

**Table 2.** Mass loss rate of HD 209458b at various physical and simulation conditions, deviating from the standard ones adopted in this paper (He/H=0.05, $F_{XUV}$=6 erg s$^{-1}$ cm$^{-2}$ at 1 a.u., $P_{base}$=0.03, $T_{base}$=1250 K, simulation domain size $10R_p$, spatial resolution close to the inner boundary r=$R_p$ $\Delta r=R_p/400$, corresponding column marked by *).

| Conditions | $\dot{M}$ ($10^{10}$ g/s) |
|---|---|
| He/H=0.01 | 6.52 |
| He/H =0.05 * | 7.26 |
| He/H =0.1 | 7.29 |
| He/H =0.2 | 6.55 |
| He/H =0.4 | 4.62 |
| $F_{XUV}$=3 erg s$^{-1}$ cm$^{-2}$ | 3.60 |
| $F_{XUV}$=12 erg s$^{-1}$ cm$^{-2}$ | 14.08 |
| zero cooling by $H_3^+$ | 8.14 |
| $P_{base}$ =0.01 bar | 6.73 |
| $P_{base}$ =0.09 bar | 7.30 |
| $T_{base}$ =1000 K | 7.00 |
| $T_{base}$ =1500 K | 7.75 |
| simulation domain r=$20R_p$ | 7.66 |
| $\Delta r=R_p/200$ | 7.68 |

In Table 2 we present the mass loss rates of HD209458b, calculated for different physical conditions and parameters of numerical modeling. The strongest influence is produced by XUV intensity, to which $\dot{M}$ is almost proportional. This is in good accordance with the energy limited explanation of the escaping PW formation (Lammer et al. 2003). We note that the largest part of XUV, contained in far FUV region, is absorbed above the $H_2$ dissociation and $H_3^+$ cooling region and goes directly to $H$ ionization and plasma heating. The XUV energy input at the temperature maximum is balanced mostly by the adiabatic cooling of the escaping material outflow. This was verified and confirmed with many previous numerical studies, which prove the relation $\dot{M} \sim F_{XUV}$. The account of infrared cooling by $H_3^+$ reduces the mass loss by about 12%. The base temperature of atmosphere, $T_{base}$, varied in the case of HD 209458b within the range of its possible values 1000–1500 K, affects the $\dot{M}$ only by 10%. The dependence of $\dot{M}$ on the base (i.e., inner boundary) atmospheric pressure (or rather density, if the temperature is fixed) should be also small, if most of the XUV energy is absorbed in the simulation box (Shaikhislamov et. all 2014). This is the reason why we impose relatively high pressures at the inner boundary, as compared to other works. One can see in Table 2, that an order of magnitude change in $P_{base}$ results in only 8% change of $\dot{M}$. This change comes from small variation of effective planet size with $P_{base}$. Moreover, the difference between $P_{base}=0.09$ and 0.03 bar is negligible in mass loss and density profiles far from the planet. On the other hand, decreasing pressure to $P_{base}=0.01$ results in a perceptible change. Variation due to other numerical factors is less than 8%. Taking into account the numerical effects and the existing uncertainties of chemical rates and $H_3^+$ emissivity, it can be stated, that the mass loss rate is calculated in our model with an accuracy of at least 15%.

**4 GJ 436b**

distances of about $20R_p$. This is in agreement with the results of 1D modeling by *Loyd et al.* (*2017*). However, in metric units the plasmasphere of GJ 436b is twice smaller than that of HD 209458b, whereas in terms of the orbital radii, both are approximately of the same scale, and in units of a host star sizes, the plasmasphere of GJ 436b is even larger. The nearby plasma environment of GJ 436b is less affected by the Coriolis and tidal forces than that in the case of HD 209458b. The details of the structure of the simulated plasmasphere of GJ 436b are given in the graphs in Figure 6.

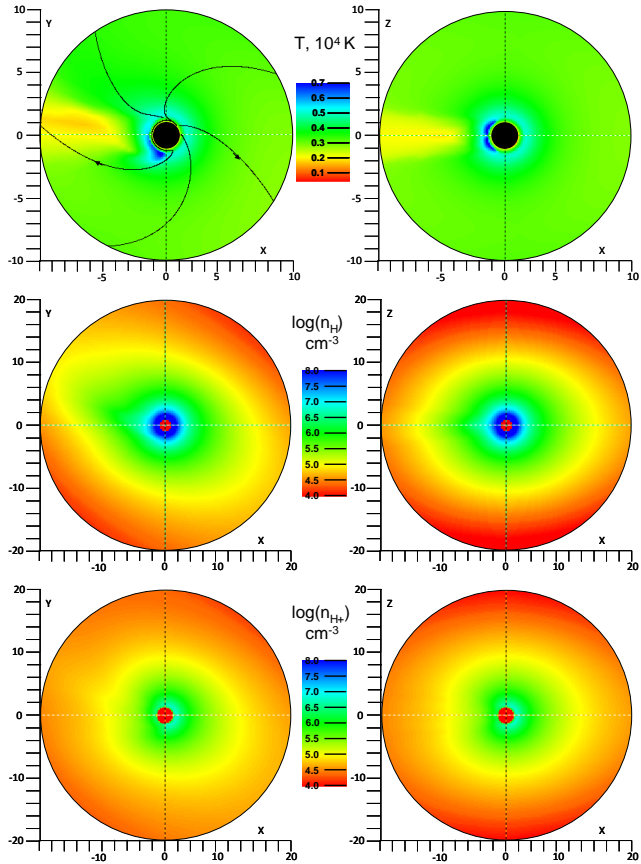

**Figure 5**. Distribution of temperature (upper row) and density (middle row) of hydrogen atoms, as well as proton density (bottom row) in the equatorial, X-Y (left column), and the meridional, X-Z (right column), planes, respectively, calculated for GJ 436b with $F_{XUV}=0.85$ erg s$^{-1}$ cm$^{-2}$ at 1 AU, $T_{base}=750$ K, $P_{base}=0.03$ bar, $He/H=0.1$. The distance is scaled in units of GJ 436b radius, $R_p$. The star is located on the right at $X=140R_p$. The material velocity streamlines are shown in the first panel. Note that the upper row panels are bounded by $r=10R_p$, whereas other panels cover twice larger area.

Figure 5 demonstrates the spatial structure of the nearby plasma environment around the warm Neptune GJ 436b, calculated for $F_{XUV}=0.85$ erg s$^{-1}$ cm$^{-2}$ at 1 AU, $T_{base}=750$ K, $P_{base}=0.03$ bar, $He/H=0.1$, which we will consider further as standard ones for this planet. In terms of the planetary radii, the plasmasphere of GJ 436b is rather large, and extends (in our simulation) up to the

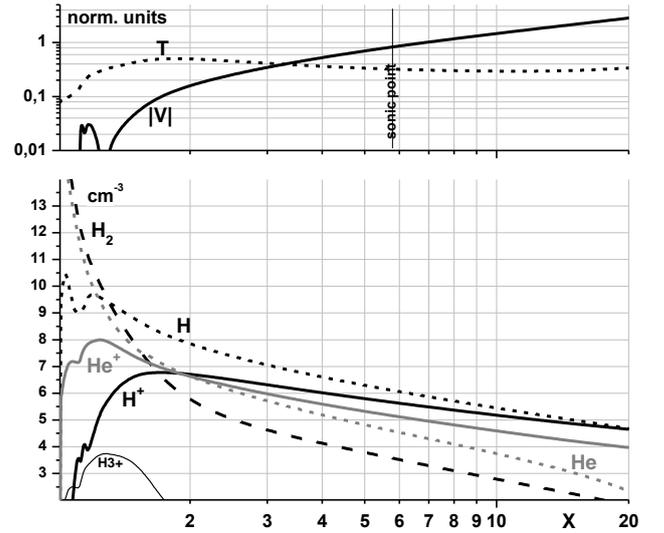

**Figure 6A.** Profiles of temperature and absolute velocity (upper panels) and density of various components (bottom panels, log scale, in cm$^{-3}$, dashed black – $H_2$, dotted black – $H$, solid black – $H^+$, dashed grey – $He$, solid grey – $He^+$, thin black – $H_3^+$) along the planet-star line. Parameters of simulations are the same as in Fig.5.

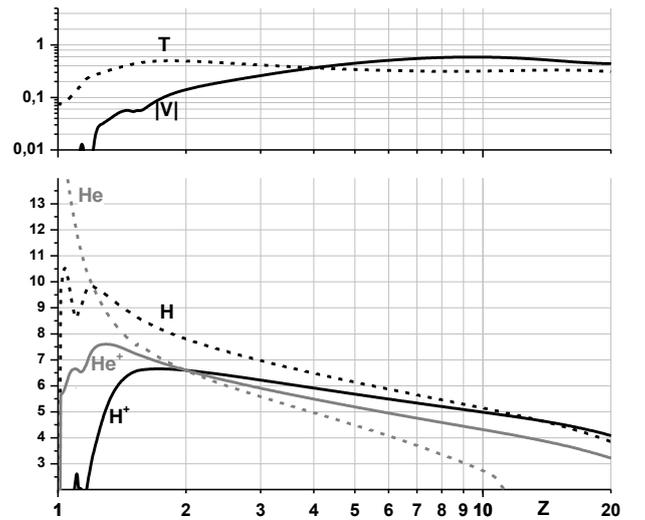

**Figure 6B.** Profiles of temperature and absolute velocity (upper panels) and density of various components (bottom panels, log scale, in cm$^{-3}$, dotted black – $H$, solid black – $H^+$, dashed grey – $He$, solid grey – $He^+$) along Z-axis. Parameters of simulations are the same as in Fig.5.

A distinctive feature of the close plasma environment around GJ 436b is that its maximum temperature does not exceed 5000 K. This is a direct consequence of low XUV flux. Because of the low temperature, the neutral $H_2$ doesn't fully dissociate, even far from the planet. Note that the region of half-ionized hydrogen is located at distances beyond $10R_p$. These and other features obtained in our 3D modeling are in a very good quantitative agreement with aeronomy simulation of (*Loyd et al. 2017*) which was only 1D, but has more involved chemistry. In this work the temperature maximum is equal to 4200 K and is positioned at $1.85R_p$, $H_2$ dissociation front is at $1.35R_p$, an outflow velocity reaches a value of 10 km/s at $8R_p$, and the mass loss rate is $\sim 0.3 \cdot 10^{10}$ g/s.

**Table 3.** Mass loss rate of GJ 436b at various physical and simulation conditions, deviating from the standard ones adopted in this paper ($He/H$=0.1, $F_{XUV}$=0.85 erg s$^{-1}$ cm$^{-2}$ at 1 a.u., $P_{base}$=0.03 bar, $T_{base}$=750 K, simulation domain size $R$=20$R_p$, spatial resolution at $r=R_p$ $\Delta r=R_p/200$, corresponding column marked by *).

| Conditions | $\dot{M}$ ($10^{10}$ g/s) |
|---|---|
| $He/H$=0.01 | 0.31 |
| $He/H$=0.05 | 0.38 |
| $He/H$=0.1 * | 0.40 |
| $He/H$=0.2 | 0.41 |
| $He/H$=0.4 | 0.42 |
| $He/H$=0.8 | 0.46 |
| $He/H$=2 | 0.53 |
| $F_{XUV}$=0.425 erg s$^{-1}$ cm$^{-2}$ | 0.19 |
| $F_{XUV}$=1.7 erg s$^{-1}$ cm$^{-2}$ | 0.88 |
| zero cooling by $H_3^+$ | 0.48 |
| $P_{base}$=0.01 bar | 0.38 |
| $P_{base}$=0.09 bar | 0.43 |
| $T_{base}$=500 K | 0.36 |
| $\Delta r=R_p/400$ | 0.39 |

In Table 3 we present the mass loss rates of GJ 436b, calculated for different physical conditions and parameters of numerical modeling. Because of the relatively low XUV flux and correspondingly large photo-ionization time of hydrogen, the quasi-stationary state in the simulations of GJ 436b is reached in about 250-300 characteristic time units $R_p/V_o$. As expected, the strongest influence on $\dot{M}$ is produced by XUV intensity. The infrared cooling by $H_3^+$ reduces the mass loss by about 20%. The base temperature of atmosphere $T_{base}$, varied in this case within the range of its possible values 500–750 K, affects the $\dot{M}$ only by 10%. An order of magnitude variation of $P_{base}$ results in only 13% mass loss change.

To illustrate the influence of $He$ abundance on the escaping atmosphere conditions in the vicinity of GJ 436b we present in Figure 7 the density and temperature profiles of molecular hydrogen (similarly to Figure 4). The density of $H_2$ (the main component in the base atmosphere) decreases with height faster for the higher $He$ abundances. This is because the presence of $He$ decreases the mean barometric gravitational scale height.

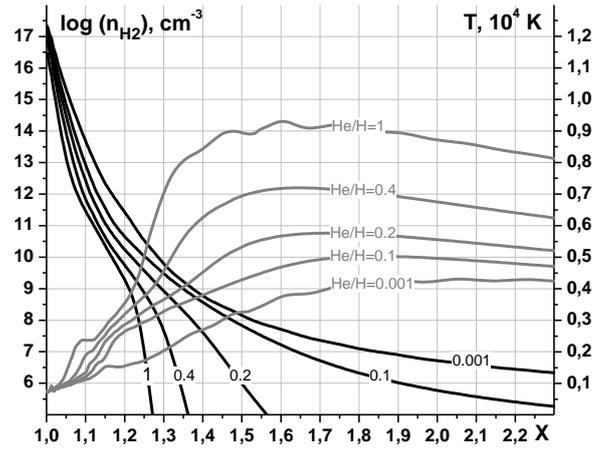

**Figure 7.** Density (black lines, left axis) and temperature (grey lines, right axis) profiles of molecular hydrogen along the planet-star line in the plasmasphere of GJ 436b, calculated for different helium abundances (indicated in the plots).

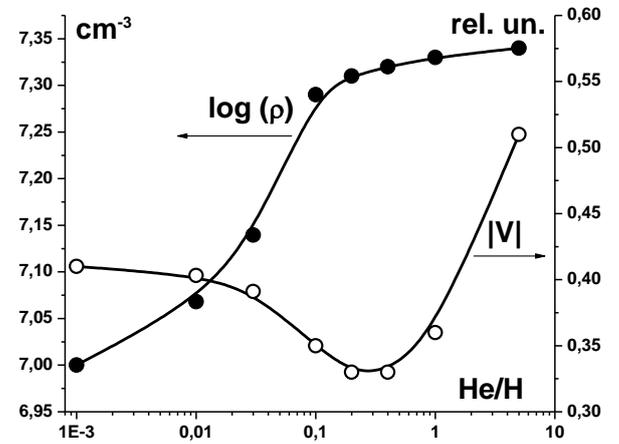

**Figure 8.** The total mass density $\rho = \sum m_i n_i$ (black circles, left axis) and PW velocity along the planet-star line (open circles, right axis) at $X=3R_p$ as a function of helium abundance, calculated for GJ 436b. Note that $\rho$ is calculated using $m_i$ in atomic units.

Faster decrease of $H_2$ with height results in less cooling due to dissociation and infrared emission by $H_3^+$ and, therefore, in higher temperatures of thermosphere. Despite of faster decrease of $H_2$ density with height, the total mass density $\rho = \sum m_i n_i$ actually increases with the increasing helium content, mainly because of helium mass and correspondingly larger mean molecular weight. This is specifically shown in Figure 8. At moderate $He$ abundances the velocity of escaping PW decreases as expected. However, the increase in total mass density compensates it, resulting in the overall increase of the mass loss rate $\dot{M}$ with the increasing $He$ abundance. When helium becomes the mass dominating component in atmosphere, one can see in Figure 8 a sharp increase of the PW velocity. The increase of the PW velocity in this case is due to larger temperature and corresponding thermal pressure gradient at higher $He$ abundances (see in Figure 7). Another feature revealed in the modeling of GJ 436b, which is qualitatively similar to that observed in the PW simulations on HD 209458b, consists in the fact that the $H_2$ envelope, extending at small $He$ abundances over a wide region around the planet, for $He/H \geq 0.2$ becomes restricted by a sharp dissociation front at heights below $1.6R_p$. Therefore, we see that helium affects the thermosphere via a complex interplay of such factors, as mean molecular mass, scale height, dissociation of $H_2$ and infrared cooling by $H_3^+$.

The dependence of mass loss on $He$ abundance is presented in Figure 9 for both exoplanets. One can see that up to moderate abundances $He/H=0.05$–$0.1$ the mass loss gradually increases by about 40%. As was stated before, this increase is related to increase of total mass density or mean molecular mass $\overline{m}_i$. However, for larger abundances $He/H >0.1$ the mass loss of HD 209458b decreases while of GJ 436b continues to rise. In case of more massive HD 209458b the decrease is explained by faster atmosphere fall off due to the same mean molecular mass (see Fig. 4). For lighter GJ 436b the faster decline of atmosphere with increasing $He$ abundance is overcompensated by increase of temperature and outflow velocity (Fig. 7 and 8). Note that for HD 209458b the temperature also increases with $He$ abundance, but the effect is more moderate.

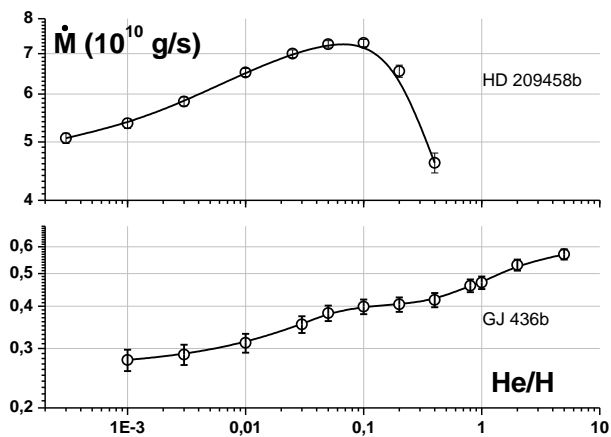

**Figure 9.** Dependence of mass loss on $He$ abundance for HD 209458b (upper panel) and GJ 436b (lower panel).

## 5 DISCUSSION AND CONCLUSION

We present a self-consistent 3D model of the expanding PW for two exoplanets: HD209458b and GJ 436b. An important and essential 3D effect quantified with the model for the planetary thermospheres consists in the redistribution of XUV heating by the zonal flows around a planet. In case of HD 209458b the zonal flows extend up to a height of about $1.25R_p$ and reach velocities of up to 1 km/s. For GJ 436b, it is several times less because of the lower XUV flux. The location of temperature maximum in the surrounding plasma is shifted for both modeled exoplanets from the sub-stellar region to the side because of the flow distortion by the Coriolis force. It significantly influences the flow direction at heights of about $5R_p$ and the created plasma envelopes are inclined in general at about 25–45 degrees relative to the planet-star line. This fact should definitely affect the Doppler broadening absorption of the spectral lines measured during the transits of exoplanets. Another specific 3D effect, verified in our simulations, is the compression of the plasmasphere towards the ecliptic plane by the stellar gravity, which in the polar directions of the tidally locked rotating system is not compensated by the centrifugal force. For HD 209458b this effect results in the formation of sharp plasmasphere boundaries at a distance of about $8R_p$ below and above the ecliptic plane.

While the total influence of the $He$ abundance on the mass loss is relatively moderate, the presence of $He$ significantly affects the structuring of the plasmasphere close to the planet. Because of a higher mass, it decreases the mean scale height of the atmosphere resulting in a faster density fall off. This lowers the $H_2$ dissociation front and for GJ 436b we see two qualitatively different thermospheres, one with an extended molecular hydrogen envelope, and another, at $He/H >0.1$ with a restricted $H_2$ area localized close to the planet. For both studied exoplanets the presence of $He$ increases the maximum temperature of thermosphere. When $He$ content is comparable to that of $H$, the temperature increases by 1.5–2 times, due to reduction of such cooling processes as $H_2$ dissociation, $H_3^+$ infrared emission and adiabatic expansion.

As has been stated before, the dependence of the scale height on the mean molecular mass is assumed in the model only as the initial state of the atmosphere. In calculations the distribution of different species is self-consistently and independently adjusted correspondingly to acting forces. When the collision length between species is much smaller than the scale height, which is the case in a dense atmosphere close to planet, they are strongly coupled and move as a single fluid. Thus, radial expansion of atmosphere transports the mixing ratio of elements from the low atmosphere upward and, if bulk transport exceeds molecular diffusion, makes it uniform everywhere. Wherever the mixing ratio is uniform, the balance between main forces – the gravity and the thermal pressure, leads to barometric equilibrium with scale height depending on mean molecular mass. However, when collision length becomes larger than the scale height the gravitational separation of elements takes place because it acts on the outflowing heavier particles stronger. In this case the abundance of heavier elements decrease with height much faster than of the lighter ones. Exactly this can be seen in Fig. 6B where at relatively rarified PW the neutral Helium becomes collisionally uncoupled from other components and its density drops off at about $10R_p$. At the same time Hydrogen and ionized Helium coupled to protons, expand to larger distances. Nevertheless, the gravitational differentiation of species has aspects which are not tackled in the present model and which can affect the actual atmosphere even close to the planet. For example, given very large time for equilibrium relaxation and relatively low hydrodynamic escape, significant diffusive separation of Helium from Hydrogen can evolve due to molecular diffusion (*Hu et al. 2015*). On the other hand, small scale turbulence and convection which is bound to be present in heated atmosphere of hot exoplanets should lead to very efficient mixing of elements ensuring constant mixing ratio.

The presence of *He* also significantly affects the photo-ionization of hydrogen close to the planet. As can be seen in Fig. 6, in case of GJ 436b most of electrons up to the height of $2R_p$ are supplied by *He* photo-ionization. When *He* abundance is negligible, the proton density maximum with a value of $3 \cdot 10^7$ cm$^{-3}$ takes place already at about $1.2R_p$. Approximately the same picture is observed in simulations of HD209458b. The reason for that consists in the fact that helium is quickly ionized by high energy photons with λ<50 nm which easily penetrate into deep layers of the thermosphere, because their absorption by hydrogen has an order of magnitude smaller cross section.

In spite of the fact that HD209458b has been simulated previously with other 3D codes, the present model is the first one where the self-consistent expansion of planetary upper atmosphere and creation of a structured plasma envelope around the planet has been studied in full details. The main novel feature of the present 3D modeling important for the interpretation of transit observations, concerns the quantitative simulation of the structure and dynamics of the escaping PW and close planetary plasma environment shaped by the Coriolis and tidal forces. In the case of warm Neptune GJ 436b, our simulation appears in a very good agreement with the first aeronomy modeling of this exoplanet (*Loyd et al. 2017*), which was just 1D, but more chemically complex including oxygen and carbon. Nevertheless, such features as temperature maximum, outflow velocity, $H_2$ half-dissociation height, $H_2$ presence far from the planet, and the total mass loss value, are quantitatively similar, and the differences within of about 25% can be attributed to 3D effects of our model. We confirm with the 3D simulations that the exosphere of GJ 436b is very extended, indeed covering most of the star disk even when the planet transits close to the disk edge.

**Acknowledgements:**
This work was supported by grant № 18-12-00080 of the Russian Science Foundation. HL and MLK acknowledge the Austrian Science Fund (FWF) projects S11606-N16, S11607-N16 and I2939-N27 of the Austrian Science Foundation (FWF). HL acknowledge also support from the FWF project P25256-N27 'Characterizing Stellar and Exoplanetary Environments via Modeling of Lyman-α Transit Observations of Hot Jupiters'. Parallel computing simulations, key for this study, have been performed at Computation Center of Novosibirsk State University, SB RAS Siberian Supercomputer Center, and Supercomputing Center of the Lomonosov Moscow State University.